\documentclass[11pt,letterpaper]{article}

\usepackage[T1]{fontenc}
\usepackage[cp1251]{inputenc}
\usepackage{textcomp}
\usepackage[centertags]{amsmath}
\usepackage{amsfonts}
\usepackage{amssymb}
\usepackage[numbers,sort&compress]{natbib}
\usepackage[pdftex]{hyperref}
\usepackage{graphicx}
\usepackage{graphbox}
\usepackage{paperinitial}

\paperinitialization{15mm}{15mm}{20mm}{10mm}{2pt}{10pt}

\DeclareMathOperator{\re}{Re}
\DeclareMathOperator{\im}{Im}
\DeclareMathOperator{\sgn}{sgn}
\DeclareMathOperator{\Ai}{Ai}
\DeclareMathOperator{\Bi}{Bi}

\newcommand{\e}{\varepsilon}
\newcommand{\vf}{\varphi}
\newcommand{\s}{\sigma}

\newcommand{\al}{\alpha}
\newcommand{\be}{\beta}
\newcommand{\ga}{\gamma}
\newcommand{\Ga}{\Gamma}

\newcommand{\la}{\lambda}

\newcommand{\ups}{\upsilon}

\newcommand{\spk}{\mathbf{k}}

\begin{document}
\frenchspacing
\allowdisplaybreaks[4]

\title{{\Large \textbf{Properties of an ultrarelativistic charged particle radiation in a constant homogeneous crossed electromagnetic field}}}

\date{}

\author{O.V. Bogdanov${}^{1),2)}$\thanks{E-mail: \texttt{bov@tpu.ru}},\; P.O. Kazinski${}^{1),2)}$\thanks{E-mail: \texttt{kpo@phys.tsu.ru}},\; and G.Yu. Lazarenko${}^{1)}$\thanks{E-mail: \texttt{lazarenko.georgijj@icloud.com}}\\[0.5em]
{\normalsize ${}^{1)}$ Physics Faculty, Tomsk State University, Tomsk 634050, Russia}\\
{\normalsize ${}^{2)}$ Department of Higher Mathematics and Mathematical Physics,}\\
{\normalsize Tomsk Polytechnic University, Tomsk 634050, Russia}}

\maketitle

\begin{abstract}

The properties of radiation created by a classical ultrarelativistic scalar charged particle in a constant homogeneous crossed electromagnetic field are described both analytically and numerically with radiation reaction taken into account in the form of the Landau-Lifshitz equation. The total radiation naturally falls into two parts: the radiation formed at the entrance point of a particle into the crossed field (the synchrotron entrance radiation), and the radiation coming from the late-time asymptotics of a particle motion (the de-excited radiation). The synchrotron entrance radiation resembles, although does not coincide with, the ultrarelativistic limit of the synchrotron radiation: its distribution over energies and angles possesses almost the same properties. The de-excited radiation is soft, not concentrated in the plane of motion of a charged particle, and almost completely circularly polarized. The photon energy delivering the maximum to its spectral angular distribution decreases with increasing the initial energy of a charged particle, while the maximum value of this distribution remains the same at the fixed observation angle. The ultraviolet and infrared asymptotics of the total radiation are also described.

\end{abstract}

\section{Introduction}

The form of radiation of an ultrarelativistic charged particle moving in a crossed electromagnetic field is discussed in many textbooks and papers (see, e.g., \cite{Ritus.2,BaiKat1,BaKaStr,BaKaStrbook}). However, according to these approaches to the problem, the radiation reaction acting on a charged particle is completely neglected or taken into account only as a small perturbation. This is not always a justified approximation, especially when the external electromagnetic field is strong enough or the particle spends a sufficiently long time in the field \cite{KazShiplde,KazAnn,BogKaz}. We describe (to our knowledge, for the first time) in detail the properties of radiation created by a classical charged scalar particle entering a constant homogeneous crossed electromagnetic field and moving in it for an infinite time with radiation reaction taken into account.

Our investigation shows that, loosely speaking, this radiation consists of two parts: the first one is an intense hard radiation created by a charged particle at the entrance point (we shall call it, for brevity, the synchrotron entrance radiation), and the second one is a comparatively weak soft radiation formed on the late-time asymptotics of particle's trajectory (the de-excited radiation) \cite{KazAnn}. The properties of the synchrotron entrance radiation resemble the properties of an ultraviolet asymptotics of the synchrotron radiation. In particular, the maximum of the spectral density is reached at the photon energy the same as for the synchrotron radiation, and the most part of the radiation is concentrated in the cone opening of the order $2\ga^{-1}$, where $\ga$ is the Lorentz factor. It is this radiation which is subject to considerable quantum corrections at high external field intensities. Notice that this is not the so-called edge radiation (see the definition, e.g., in \cite{EdRadTh,SynchRad2015}). The de-excited radiation is soft, its spectrum stretches from far-infrared to hard X-rays (for $\ga\sim10^4$ and the field strengths corresponding to the intensity $I\sim 10^{22}$ W/cm${}^2$), and the quantum corrections to it are negligible. We find the region of observation angles and energies of photons where these two types of radiation can be discerned and observed.

Besides that the synchrotron entrance and de-excited radiations are distinguished by the ranges of the photon energies and observation angles where these radiations are concentrated, it turns out that they possess distinct polarization properties. The de-excited radiation is almost completely circularly polarized, while the synchrotron entrance radiation has mostly a linear polarization. We present a detailed description of the polarization properties of these two types of radiations.

The study of properties of radiation of an ultrarelativistic charged particle in a crossed electromagnetic field is important for two reasons, at least. First, in the ultrarelativistic limit, every electromagnetic field becomes crossed in the momentary comoving frame. So we may expect that, in a certain approximation, the properties of radiation we discussed in this paper should be inherent to any electromagnetic field, which can be considered as constant and homogeneous on the radiation formation length scale. Second, the crossed field approximation is the standard one \cite{Ritus.2,PiMuHaKermp.2} in considering the radiation of charged particles in strong laser fields. The intensity of the laser radiation, which will become accessible in the nearest future \cite{ELI.2,ECELS.2}, allows one to observe the radiation reaction effects we discuss \cite{KazShiplde,KazAnn,BogKaz,PiHaKei1,MaPiKei,PiHaKei2,PiMuHaKermp.2}. The study of classical radiation is relevant even in that domain of parameters where one expects considerable quantum corrections. In this case, the classical results can be used to distinguish clearly the quantum corrections. This idea was employed, for example, in a recent proposal \cite{HGIM}. Notice that the numerical simulations of electron's dynamics and its radiation in a strong laser wave and in crystals with radiation reaction taken into account already present in the literature (see, e.g., \cite{PiHaKei1,MaPiKei,PiHaKei2,SchlTikh.2,PiWiUg,HHM1,OTMT,YNMJ,VGFS,HHM2}). However, the problem of radiation of electrons in a crossed field has somehow escaped the scope of these papers.

\section{Notation and solution to the equations of motion}

The action functional of a charged particle with the charge $e$ and the mass $m$ interacting with the electromagnetic field $A_\mu$ on the Minkowski spacetime $\mathbb{R}^{1,3}$ with the metric $\eta_{\mu\nu}=diag(1,-1,-1,-1)$ has the form
\begin{equation}\label{action particl}
    S[x(\tau),A(x)]=-m\int{d\tau\sqrt{\dot{x}^2}}-e\int{d\tau A_\mu
    \dot{x}^\mu}-\frac1{16\pi}\int{d^4xF_{\mu\nu}F^{\mu\nu}},
\end{equation}
where $F_{\mu\nu}:=\partial_{[\mu}A_{\nu]}$ is the strength tensor of the electromagnetic field, and the speed of light $c=1$.

The Landau-Lifshitz equation in the natural parametrization $\dot{x}^2=1$ is written as \cite{LandLifshCTF.2}
\begin{equation}\label{lde_ini}
    m\ddot{x}_\mu =eF_{\mu\nu}\dot{x}^\nu+\frac23e^2\Big(\frac{e}{m}\dot{F}_{\mu\nu}\dot{x}^\nu+\frac{e^2}{m^2}F_{\mu\nu}F^{\nu\rho}\dot{x}_\rho- \frac{e^2}{m^2}\dot{x}^\la F_{\la\nu}F^{\nu\rho}\dot{x}_\rho\dot{x}_\mu\Big).
\end{equation}
For the thorough discussion of its applicability to the problem at hand see, e.g., \cite{KazAnn,BogKaz} and references therein. Let us choose the Compton wavelength $l_C:=\hbar/mc$ as the length unit:
\begin{equation}
    x^\mu\rightarrow l_C x^\mu,\qquad \tau\rightarrow l_C\tau.
\end{equation}
Then Eq. \eqref{lde_ini} is reduced to
\begin{equation}
    \ddot{x}_\mu =f_{\mu\nu}\dot{x}^\nu+\la(\dot{f}_{\mu\nu}\dot{x}^\nu+f_{\mu\nu}f^{\nu\rho}\dot{x}_\rho- \dot{x}^\la f_{\la\nu}f^{\nu\rho}\dot{x}_\rho\dot{x}_\mu),
\end{equation}
where $x^\mu$, $\tau$, $f_{\mu\nu}$, and $\la$ are dimensionless quantities, and $f_{\mu\nu}=\sgn(e)F_{\mu\nu}/E_0$. It is useful to bear in mind that the electromagnetic field strength is measured in the units of the critical field $E_0$, and the unit of energy is the electron rest energy:
\begin{equation}
\begin{gathered}
    l_C\approx3.86\times 10^{-11}\;\text{cm},\qquad t_C\approx1.29\times 10^{-21}\;\text{s},\qquad m\approx 5.11\times10^5\;\text{eV},\\
    E_0=\frac{m^2}{|e|\hbar}\approx 4.41\times 10^{13}\;\text{G}=1.32\times 10^{16}\;\text{V/cm},\qquad\la=\frac{2\al}3\approx\frac{2}{411},
\end{gathered}
\end{equation}
The modern accelerator facilities are able to accelerate electrons up to $20$ GeV and higher, and the intensities of the achievable at the present moment laser fields \cite{laser_tod} are of the order $10^{22}$ W/cm${}^2$. These data correspond to
\begin{equation}\label{experiment}
    \ga\approx10^5,\qquad \omega\approx1.47\times10^{-4},\qquad \la\omega\approx7.14\times10^{-7},
\end{equation}
where $\omega$ is the electromagnetic field strength in the laser wave.

The field strength for the constant homogeneous crossed electromagnetic field reads as
\begin{equation}\label{fmunu_cross}
    f^{\mu\nu}=\omega e_-^{[\mu}e_1^{\nu]},\qquad f^2_{\mu\nu}=\omega^2e^-_\mu e^-_\nu,\qquad E_x=-\omega,\quad H_z=\omega,\qquad x\geq0,
\end{equation}
where $\omega$ is a constant, $x_-=x^0-x^2$, and the $4$-vectors $e_-^\mu=(1,0,1,0)$, $e_1^\mu=(0,1,0,0)$ were introduced. Any $4$-vector can be represented in the form
\begin{equation}
    j_\mu=\frac12(j_+ e_\mu^-+j_- e^+_\mu)-j_1 e^1_\mu-j_3 e^3_\mu,
\end{equation}
where $e_+^\mu=(1,0,-1,0)$, $e_3^\mu=(0,0,0,1)$, and $j_a:=e_a^\mu j_\mu$. In particular,
\begin{equation}\label{scalar_prod}
    \ups_-\ups_+-\ups^2_1-\ups_3^2=1,\qquad j_\mu^*j^\mu=\re(j_+j_-^*)-|j_1|^2-|j_3|^2, \quad j_\mu\in \mathbb{C},
\end{equation}
where $\ups_a:=\dot{x}_a$, and the dot denotes the derivative with respect to the natural parameter. The solution to the Landau-Lifshitz equation for such a field configuration can be cast into the form (see \cite{KHRL.2} and also \cite{NikishLLsol.2,Piazza.2,HLREKR.2,KazAnn})
\begin{equation}\label{moment_comp_cros}
\begin{gathered}
    r_+=r_3^2(0)+\Big[\bar{r}(0)+\frac1{2\la\omega}(\ups_-^{-1}(0)+\la\omega^2x_-)^2\Big]^2+(\ups_-^{-1}(0)+\la\omega^2x_-)^2,\\ r_1=\bar{r}(0)+\frac1{2\la\omega}(\ups_-^{-1}(0)+\la\omega^2x_-)^2,\qquad \ups_-=(\ups_-^{-1}(0)+\la\omega^2x_-)^{-1},\qquad r_3=r_3(0),
\end{gathered}
\end{equation}
where $r_a:=\ups_a/\ups_-$ and $\bar{r}(0):=r_1(0)-(2\la\omega\ups^2_-(0))^{-1}$. The trajectory is written accordingly
\begin{equation}
\begin{split}
    x_+&=x_+(0)-\frac{\bar{r}(0)+\la\omega}{3\la^2\omega^3\ups_-^3(0)}-\frac{r_3^2(0)+\bar{r}^2(0)}{\la\omega^2\ups_-(0)}-\frac{1}{20\la^3\omega^4\ups_-^5(0)}+\\ &+\sqrt{\frac{2}{\la}}|\omega|^{-3/2}\Big[\big(r_3^2(0)+\bar{r}^2(0)\big)s+\frac{2}{3}\sgn(\omega)(\bar{r}(0)+\lambda\omega)s^3+\frac{s^5}{5}\Big],\\
    x_1&=x_1(0)-\frac{\bar{r}(0)}{\la\omega^2\ups_-(0)}-\frac{1}{6\la^2\omega^3\ups_-^3(0)}+\sqrt{\frac{2}{\la}}|\omega|^{-3/2}\Big[\bar{r}(0)s+\sgn(\omega)\frac{s^3}{3}\Big],\\
    x_3&=x_3(0)-\frac{r_3(0)}{\la\omega^2\ups_-(0)}+\sqrt{\frac{2}{\la}}|\omega|^{-3/2}r_3(0)s,\\
    x_-&=\sqrt{\frac{2}{\la}}|\omega|^{-3/2}s-\frac{1}{\la\omega^2\ups_-(0)},
\end{split}
\end{equation}
where the last equality is the definition of $s$, and we put $x_-(0)=0$. The velocity components \eqref{moment_comp_cros} in terms of this new variable become
\begin{equation}
    r_+=\bar{r}^2(0)+r^2_3(0)+2\sgn(\omega)(\bar{r}(0)+\lambda\omega)s^2+s^4,\qquad r_1=\bar{r}(0)+\sgn(\omega)s^2,\qquad r_3=r_3(0).
\end{equation}
We suppose that the charged particle enters the electromagnetic field at the instant $\tau=0$ and then moves in it for an infinite time.

\section{Radiation}

\paragraph{General formulas.}

The spectral angular distribution of radiation of one charged particle summed over the photon polarizations is written as
\begin{equation}\label{spectr_dens}
    d \mathcal{E}(\spk)=|\mathbf{E}(k)|^2\frac{R^2d\spk}{4\pi^2k_0^2}=-e^2 j_\mu^*(k) j^\mu(k)\frac{d\spk}{4\pi^2},\qquad k^2=0,
\end{equation}
where
\begin{equation}\label{current_fourier}
    j_\mu(k)=\int_{\tau_1}^{\tau_2}d\tau \dot{x}_\mu e^{-ik_\nu x^\nu(\tau)}-\frac{i\dot{x}_\mu}{k_\nu\dot{x}^\nu}\Big|_{\tau_1}^{\tau_2},\qquad k^\mu j_\mu(k)=0.
\end{equation}
Notice that if one calculates the contribution of the boundary points in the integral \eqref{current_fourier} by the WKB method (when this approximation is applicable) then the leading WKB contribution is canceled by the out of the integral term in \eqref{current_fourier}. The projections of the radiation electric field in the wave zone take the form
\begin{equation}
    E_\al(k)=-iek_0\frac{e^{-ik_0 R}}{R}b^\mu_{(\al)}j_\mu(k),\qquad\al=1,2,
\end{equation}
where $b^\mu_{(\al)}$ are the physical photon polarization vectors and $R$ is a distance from the source of radiation to the observation point. The Stokes parameters are
\begin{equation}\label{Stokes_par_gen}
    \xi_1=2\frac{\re(E_1E^*_2)}{|\mathbf{E}|^2},\qquad\xi_2=2\frac{\im(E_1E_2^*)}{|\mathbf{E}|^2},\qquad\xi_3=\frac{|E_1|^2-|E_2|^2}{|\mathbf{E}|^2}.
\end{equation}
In particular,
\begin{equation}\label{Stokes_circ}
    \xi_1^2+\xi_3^2=1-\xi_2^2=\frac{|\mathbf{E}^2|^2}{|\mathbf{E}|^4}=\frac{|j_\mu j^\mu|^2}{(j^*_\nu j^\nu)^2}.
\end{equation}
In the case of circularly polarized wave, $\xi_2=\pm1$. At $\xi_2=0$, the electromagnetic wave is linearly polarized and
\begin{equation}
    \xi_1=\sin(2\vartheta),\qquad\xi_3=\cos(2\vartheta),
\end{equation}
where $\vartheta$ is the angle between the polarization plane and the axis with the unit vector $\mathbf{b}_{(1)}$.

The problem is reduced to the evaluation of the integrals \eqref{current_fourier}, where, in our case, $\tau_1=0$ and $\tau_2=\infty$. The out of the integral term corresponding to $\tau_2=\infty$ is absent in \eqref{current_fourier}. It turns out that the main characteristics of the radiation created by a charged particle in the crossed field with radiation reaction taken into account can be described analytically with a rather good accuracy.

\paragraph{De-excited radiation.}

The expression standing in the exponent in \eqref{current_fourier} can be cast into the form
\begin{multline}\label{kmuxmu}
    k_\mu x^\mu= k_\mu x^\mu(0)-\frac{k_-}{2}\Big[\frac{|z|^2}{\la\omega^2\ups_-(0)}-\frac{z'-\la\omega}{3\la^2\omega^3\ups_-^3(0)}-\frac{1}{20\la^3\omega^4\ups_-^5(0)}\Big]\\ +\sqrt{\frac{2}{\la}}\frac{k_-}{10|\omega|^{3/2}}\Big[5|z|^2s-\frac{10}{3}\sgn(\omega)(z'-\la\omega)s^3+s^5\Big],
\end{multline}
where
\begin{equation}
    z\equiv z'+iz'':=\zeta'-\bar{r}(0)+i(\zeta''-r_3(0)),\qquad \zeta\equiv \zeta'+i\zeta'':=\frac{k_1+i|k_3|}{k_-},\qquad k_-=\frac{2k_0}{1+|\zeta|^2}.
\end{equation}
The coordinates $\zeta'$, $\zeta''$ are coordinates of the stereographic projection of the sphere $|\spk|=1$ to the plane $(k^1,k^3)$ from the pole $(0,1,0)$. In the ultrarelativistic limit, $r_1$ and $r_3$ are also coordinates of the stereographic projection of the sphere $|\boldsymbol{\ups}|=1$ to the plane $(\ups^1,\ups^3)$ from the pole $(0,1,0)$.

It is convenient to introduce the notation $\phi:=\phi'+i\phi''$, $\rho:=\rho'+i\rho''$, where
\begin{equation}
\begin{alignedat}{2}
    \phi'&:=\sgn(\omega)(z'-\la\omega),&\qquad |\phi|^2&:=|z|^2\;\Leftrightarrow\;(\phi'')^2=(z'')^2+2\la\omega z'-\la^2\omega^2,\\
    \rho'&:=-\sgn(\omega)(\bar{r}(0)+\la\omega),&\qquad|\rho|^2&:=\bar{r}^2(0)+r_3^2(0),\\
    \e&:=\sqrt{\frac{2}{\la}}\frac{k_-}{10|\omega|^{3/2}},
\end{alignedat}
\end{equation}
and $\psi_0$ is the expression in the first line on the right-hand side of equality \eqref{kmuxmu}. In particular, the following relation holds
\begin{equation}
    (\phi'-\rho)(\phi'-\rho^*)-(\phi'-\rho'-\la|\omega|)^2-r_3^2(0)=2\la|\omega|\phi'.
\end{equation}
Using the new notation,
\begin{equation}
    r_+=(s^2-\rho)(s^2-\rho^*),\qquad r_1=-\la\omega+\sgn(\omega)(s^2-\rho').
\end{equation}
This notation is useful for the evaluation of integrals \eqref{current_fourier}.

\begin{figure}[t]
\centering
\includegraphics*[width=0.35\linewidth]{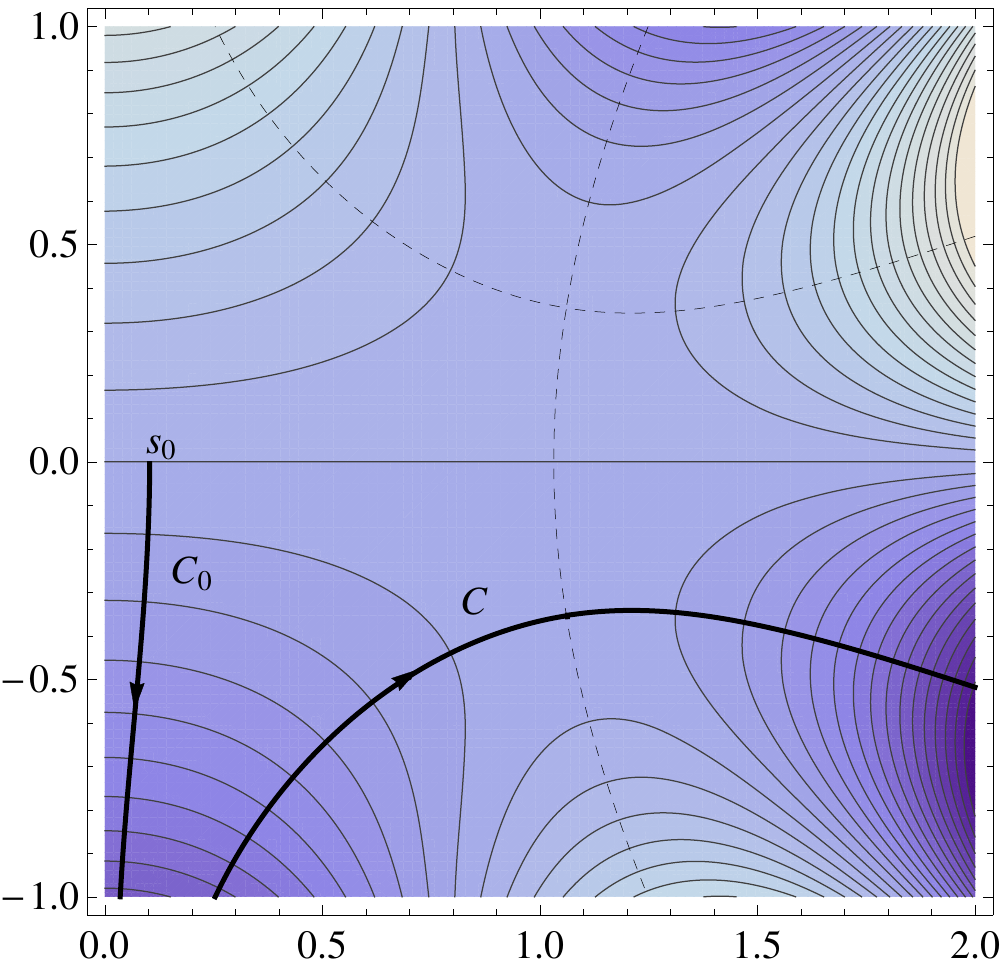}
\caption{{\footnotesize The deformation of the integration contour in the $s$ plane. The point $s_0$ corresponds to $(2\la|\omega|\ups^2_-(0))^{-1/2}$}.}
\label{cont_airy}
\end{figure}

The integrals determining the Fourier transform of $j_\mu$ are of the same type as considered in \cite{KazAnn} (see also App. \ref{F_phi_Asymptotes}) and are calculated analogously. It is useful to deform the integration contour as it is depicted in Fig. \ref{cont_airy}. The radiation created by a de-excited charged particle corresponds to the case (i) in \eqref{fphi_cases}. Let us change the integration variable
\begin{equation}\label{y_var}
    y=(20\e\phi')^{1/3}(s-\sqrt{\phi'}).
\end{equation}
Then
\begin{equation}
\begin{split}
    k_\mu x^\mu&=\psi_0+\e\big[\tfrac83(\phi')^{5/2}+5(\phi'')^2(\phi')^{1/2}\big]+By+y^3/3+hy^4/4+h^2y^5/20,\\
    h:&=(20\e)^{-1/3}(\phi')^{-5/6},\qquad B:=5\Big(\frac{\e^2}{20\phi'}\Big)^{1/3}(\phi'')^2.
\end{split}
\end{equation}
Having neglected the out of the integral terms in \eqref{current_fourier}, the contributions coming from the contour $C_0$, and assuming that \eqref{asympt_cond1} is fulfilled, we obtain up to the terms of the order $h^2$:
\begin{equation}\label{current_fourier_2}
\begin{split}
    c^{-1}_0j_+&=(\phi'-\rho)(\phi'-\rho^*)I_0+h\big[4\phi'(\phi'-\rho')I_1 -(\phi'-\rho)(\phi'-\rho^*)\frac{iI_4}{4} \big]\\
    &+h^2\Big[2\phi'(3\phi'-\rho')I_2-\phi'(\phi'-\rho')iI_5-(\phi'-\rho)(\phi'-\rho^*) \big(\frac{I_8}{32}+\frac{iI_5}{20}\big)\Big],\\
    c^{-1}_0j_-&=I_0-\frac{ih}{4}I_4-h^2\big(\frac{I_8}{32}+\frac{iI_5}{20}\big),\qquad j_3=r_3(0)j_-,\\
    c^{-1}_0j_1&=\sgn(\omega)\Big\{(\phi'-\rho'-\la|\omega|)I_0 +h\big[2\phi'I_1-(\phi'-\rho'-\la|\omega|)\frac{iI_4}{4}\big]\\ &+h^2\Big[\phi'(I_2-\frac{iI_5}{2})-(\phi'-\rho'-\la|\omega|)\big(\frac{I_8}{32}+\frac{iI_5}{20}\big) \Big]\Big\}.
\end{split}
\end{equation}
It is the order $h^2$ where the leading contribution to the de-excited radiation comes from. Here
\begin{equation}
    c_0:=e^{-i\psi_0-i\e\big[\tfrac83(\phi')^{5/2}+5(\phi'')^2(\phi')^{1/2}\big]}\sqrt{\frac{2}{\la}}|\omega|^{-3/2}(20\e\vf')^{-1/3},\qquad I_n(B):=\int_C dtt^ne^{-i(Bt+t^3/3)}.
\end{equation}
The integrals $I_n(B)$, $n\in \mathbb{N}$, are expressed through the Airy functions (see \cite{KazAnn}). As a result,
\begin{equation}\label{jjbar}
    \re(j_+j_-^*)-|j_1|^2-|j_3|^2=|c_0|^2\Big\{2\la|\omega|\phi'\Big[I_0^2+h^2\big(I_0I_2+\frac{iI_1I_4}{2}-\frac{3iI_0I_5}{5}-\frac{I_0I_8}{16}+\frac{I_4^2}{16}\big)\Big] +4h^2(\phi')^2(I_1^2+I_0I_2) \Big\}.
\end{equation}
In the case when the applicability conditions \eqref{applicability} of the approximation considered are fulfilled, the last term in the curly brackets dominates over the first terms in the square brackets.

\begin{figure}[tp]
\centering
a)\;\includegraphics*[align=c,width=0.5\linewidth]{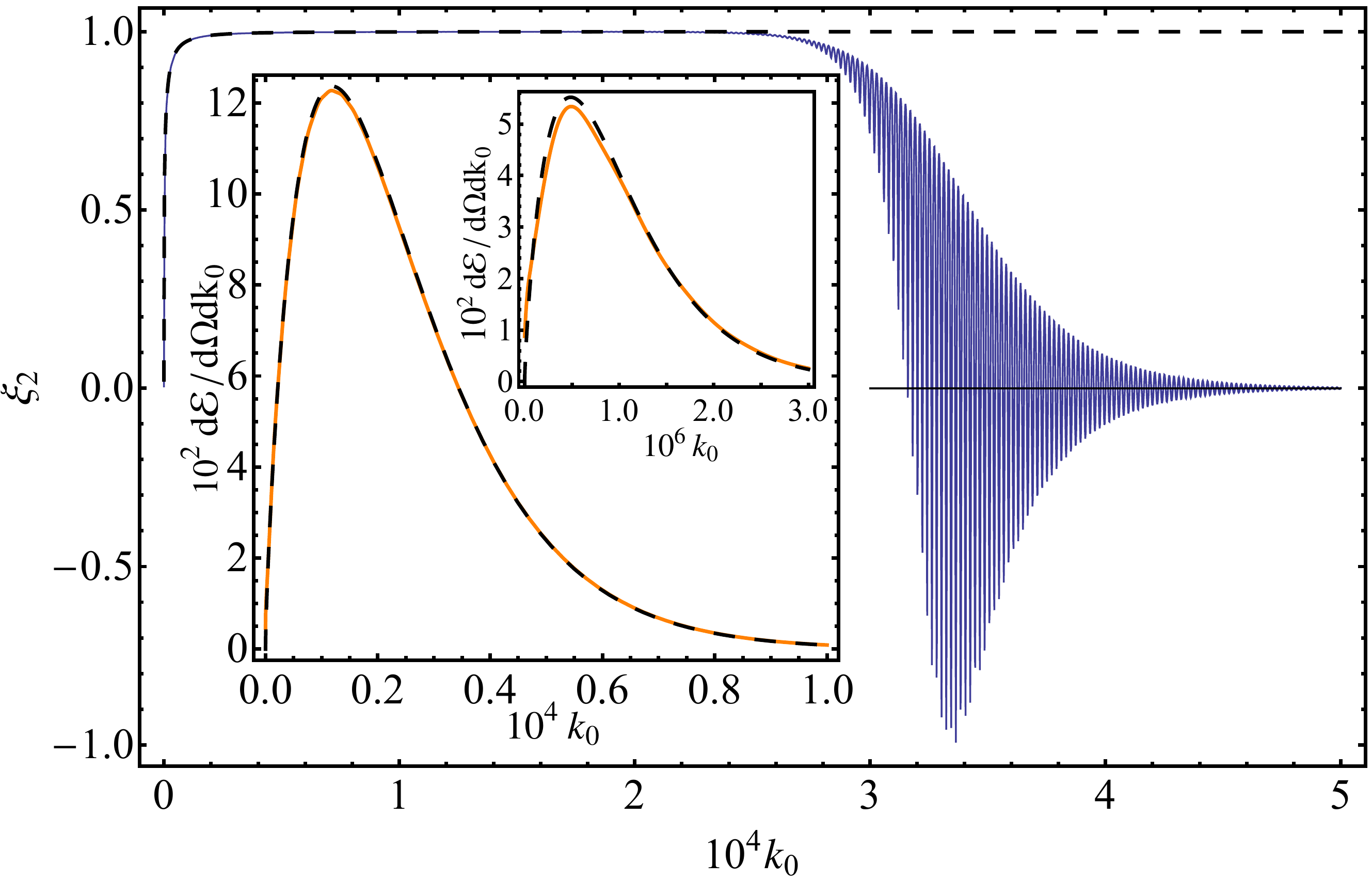}\\
b)\;\includegraphics*[align=c,width=0.32\linewidth]{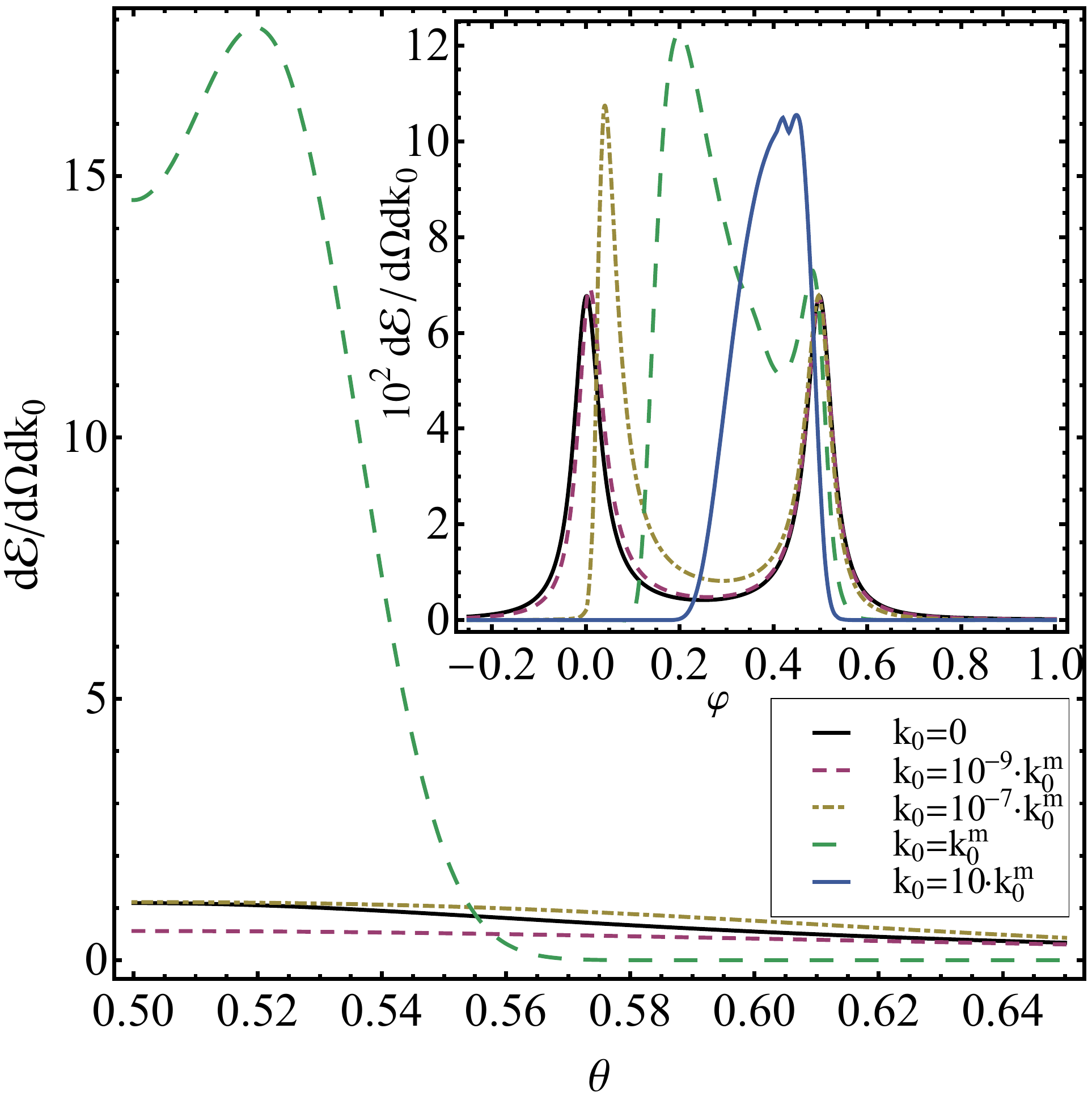}\;\;
c)\;\includegraphics*[align=c,width=0.33\linewidth]{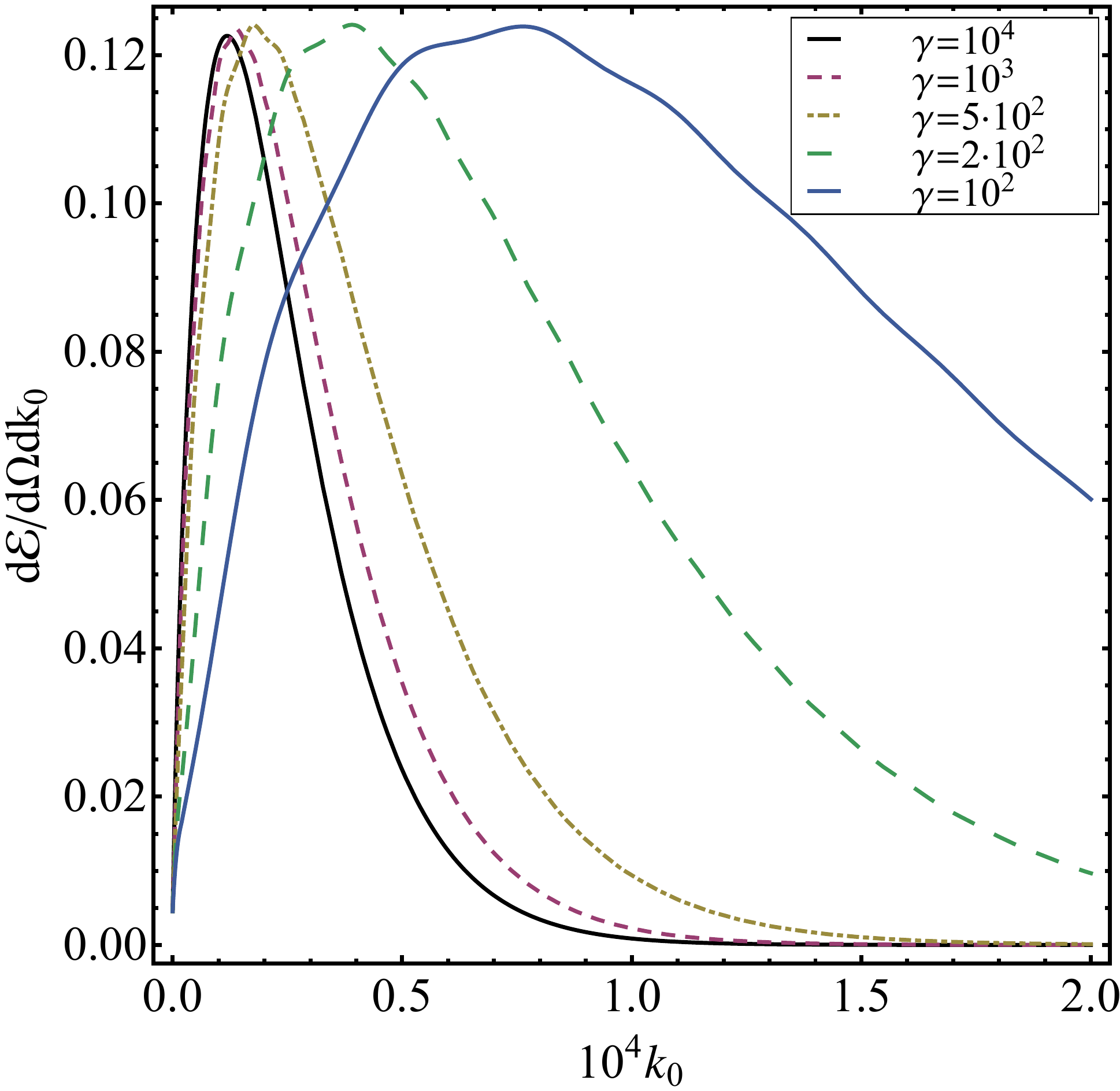}\\
d)\;\includegraphics*[align=c,width=0.32\linewidth]{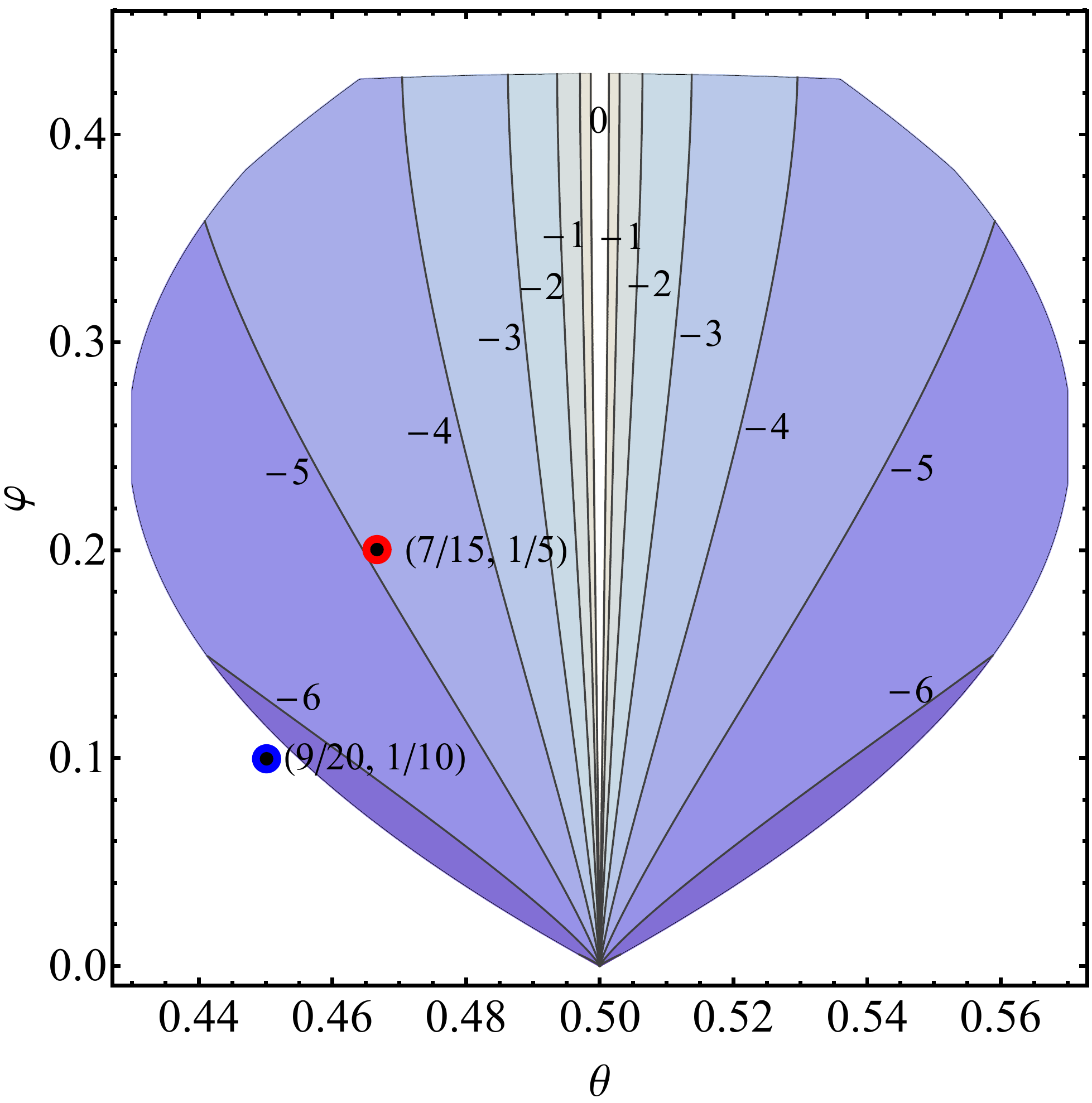}\;\;
e)\;\includegraphics*[align=c,width=0.325\linewidth]{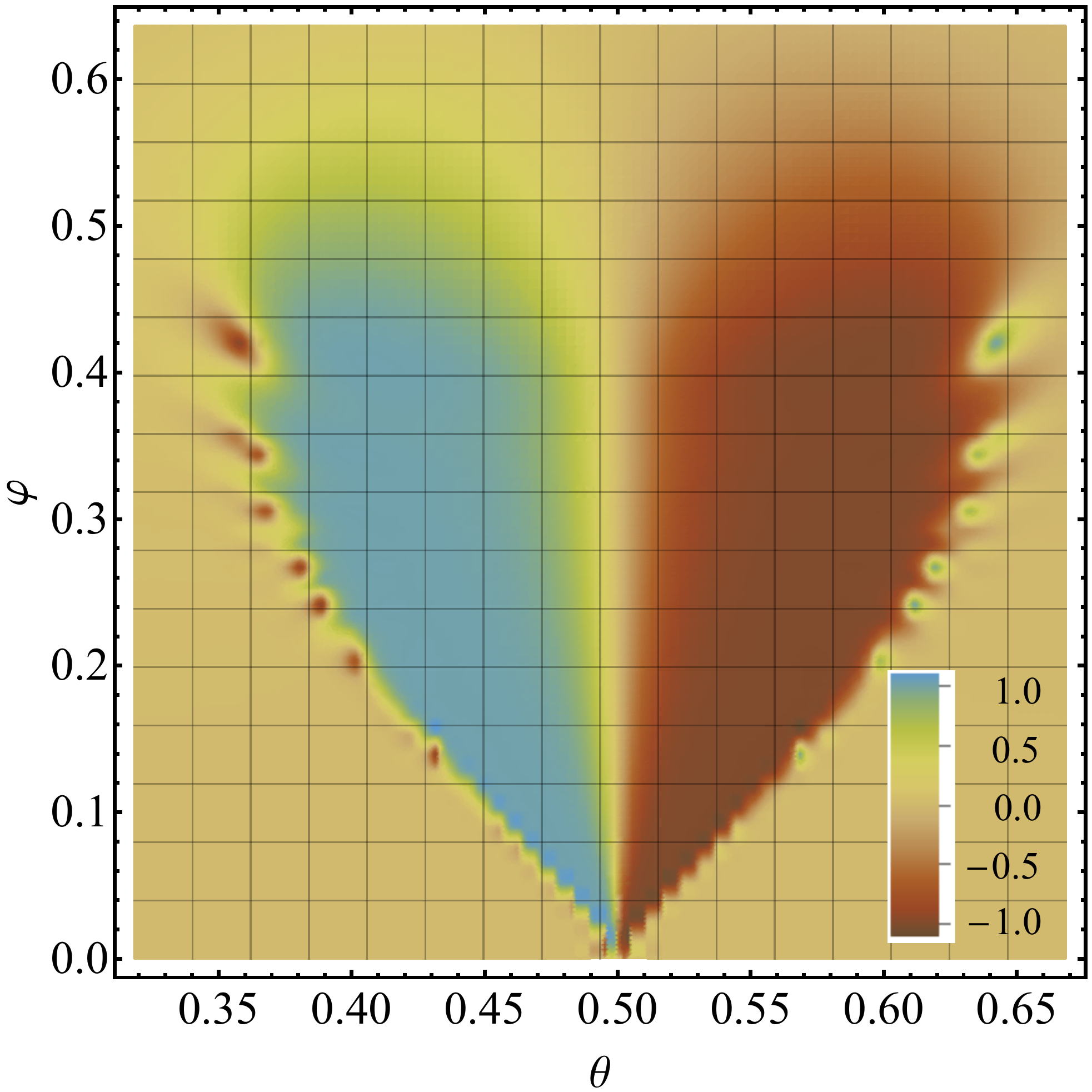}
\caption{{\footnotesize De-excited radiation for $\ga=5\times10^3$, $\omega=-1.47\times10^{-4}$, $r_3(0)=0$, and the entrance angle $\pi/2$ counted from the $x^2$ axis. a) The dependence of the Stokes parameter $\xi_2$ on the energy of radiated photon at the observation angles $\theta=7\pi/15$, $\vf=\pi/5$, where $\theta$ and $\vf$ are the polar and azimuthal angles of the spherical coordinate system. The angle $\vf$ is counted from the $x^2$ axis towards the $x^1$ axis. The blue line is the numerical result. The dashed line is \eqref{Stokes_xi2}. The de-excited radiation is almost completely circularly polarized. The radiation at extremely low and high frequencies is linearly polarized in accordance with \eqref{polariz_vect_UV}, \eqref{spectr_dens_0}. The inset: The spectral angular distribution at $\theta=7\pi/15$, $\vf=\pi/5$. The orange line is the numerical result. The dashed line is \eqref{spectr_dens_2}. The small plot is the same for $\theta=9\pi/20$, $\vf=\pi/10$. We see that formula \eqref{spectr_dens_2} may describe the radiation quite well even out of the applicability domain \eqref{applicability}. b) The spectral angular distribution for different photon energies at $\vf=\pi/5$. Here $k_0^m\approx1.20\times10^{-5}$ is given by \eqref{energy_max} at $\theta=7\pi/15$, $\vf=\pi/5$. The de-excited radiation is not concentrated in the plane of particle's motion. The inset: The same for $\theta=7\pi/15$. The two peaks correspond to the de-excited and synchrotron entrance radiations. The curve for $k_0=0$ coincides with \eqref{spectr_dens_0}. c) The dependence of the spectral angular distribution on the initial energy of a charged particle at $\theta=7\pi/15$, $\vf=\pi/5$. The maximum of the curve is well described by \eqref{spectr_dens_2}, \eqref{energy_max} and does not change in a wide range of $\ga$'s. d) The applicability domain of \eqref{spectr_dens_2} as it follows from \eqref{applicability} for the radiated photon energy taken at the maximum \eqref{energy_max}. The level lines are the lines of a constant $\lg k_0^m$. e) The Stokes parameter $\xi_2$ at the photon energy $k_0^m\approx1.20\times10^{-5}$. The plateaus $|\xi_2|\approx1$ of the circularly polarized de-excited radiation are clearly seen.}}
\label{rad2_plots}
\end{figure}

So, according to \eqref{scalar_prod}, \eqref{spectr_dens}, the spectral angular distribution of the de-excited radiation has the form (cf. \cite{KazAnn})
\begin{equation}\label{spectr_dens_2}
    d \mathcal{E}(\spk)\approx4e^2B\big\{B[\Ai(B)]^2+[\Ai'(B)]^2\big\}\frac{dk_0d\zeta' d\zeta''}{(\phi'')^2}=4e^2B\big\{B[\Ai(B)]^2+[\Ai'(B)]^2\big\}\frac{d\spk}{k_-^2(\phi'')^2}.
\end{equation}
The maximum of this radiation is reached at the photon energy
\begin{equation}\label{energy_max}
    k^m_0=(B'_{ext}|\omega|)^{3/2}(2\la)^{1/2}(1+|\zeta|^2)\frac{(\phi')^{1/2}}{|\phi''|^3},
\end{equation}
where $B'_{ext}\approx 0.8$, and it is supposed that $\phi'\geq0$. If
\begin{equation}
    |r_1(0)|\gg (2\la\omega\ups^2_-(0))^{-1},\qquad |r_1(0)|\gg \lambda|\omega|,
\end{equation}
which are satisfied as a rule, then
\begin{equation}
    \phi'\approx\sgn(\omega)(\zeta'-r_1(0)),\qquad \phi''\approx|\zeta''-r_3(0)|.
\end{equation}
In contrast to the synchrotron radiation, the right-hand member of \eqref{energy_max} depends rather weakly on the energy of the incident particle. In increasing the energy, the magnitude of $k^m_0$ slightly declines due to decreasing of $\vf'$ and tends to a finite value independent of the incident particle energy (see Fig. \ref{rad2_plots}). At that, according to the approximate formula \eqref{spectr_dens_2}, the value of the spectral density at the maximum does not virtually change.

The numerical simulations show (see Fig. \ref{rad2_plots}) that there exists a region of observation angles where formula \eqref{spectr_dens_2} provides a quite good approximation for the spectral density of radiation for all the photon energies. In this case, one can derive the angular distribution of radiation integrating \eqref{spectr_dens_2} over $k_0$. Since
\begin{equation}
    \int_0^\infty dtt^{3/2}\big\{t[\Ai(t)]^2+[\Ai'(t)]^2\big\}=\frac{1}{16},
\end{equation}
we deduce
\begin{equation}
    d \mathcal{E}(\theta,\vf)=\frac{3e^2}{32}(2\la|\omega|^3)^{1/2}(1+|\zeta|^2)^3\frac{(\phi')^{1/2}}{|\phi''|^5}d\Omega.
\end{equation}
This expression tends to infinity for $|\zeta|\rightarrow+\infty$ or $\phi''\rightarrow0$. The first case corresponds to the radiation along the $x^2$ axis. The second case refers to the observation angles lying in the plane of motion of a charged particle. In the both cases, formula \eqref{spectr_dens_2} is not valid (see the applicability domain \eqref{applicability} and Fig. \ref{rad2_plots}).

Employing \eqref{Stokes_circ} and \eqref{current_fourier_2}, it is not difficult to find the expression for the Stokes parameter $\xi_2$ of the de-excited radiation. Up to the terms of the order $h^2$, we have
\begin{equation}
    j_\mu j^\mu=c^2_0\Big\{2\la|\omega|\phi'\Big[I_0^2-\frac{ih}{2}I_0\big(I_4+4iI_1\big)+h^2\big(I_0I_2-\frac{iI_1I_4}{2}-\frac{3iI_0I_5}{5}-\frac{I_0I_8}{16}-\frac{I_4^2}{16}\big)\Big] -4h^2(\phi')^2(I_1^2-I_0I_2) \Big\}.
\end{equation}
In the parameter space we consider, the leading contribution gives the last term in the curly brackets. Then, in accordance with \eqref{Stokes_circ}, we derive
\begin{equation}\label{Stokes_xi2}
\begin{split}
    |\xi_2|\approx\sqrt{1-\frac{(B[\Ai(B)]^2-[\Ai'(B)]^2)^2}{(B[\Ai(B)]^2+[\Ai'(B)]^2)^2}} &=\Big(\frac{2}{3}\Big)^{1/3}\Ga\Big(\frac16\Big)\sqrt{\frac{B}{\pi}}-\Big(2+\frac{36\Ga^3(7/6)}{\pi^{3/2}}\Big)B^{3/2}+O(B^{5/2})\\
    &=1-\frac{1}{32B^3}+\frac{3}{64 B^{9/2}}+O(B^{-6}).
\end{split}
\end{equation}
The series for $B$ large is asymptotic. In particular, at $B=B'_{ext}$, we have
\begin{equation}\label{Stokes_max}
    |\xi_2|\approx0.98,
\end{equation}
i.e., the radiation is almost completely circularly polarized. The value \eqref{Stokes_max} of the Stokes parameter at the maximum is universal. The plot of the dependence of $\xi_2$ on the angles and the radiated photon energy is given in Fig. \ref{rad2_plots}.

One can find a rough estimate for the applicability conditions of the expressions \eqref{current_fourier_2}. These conditions define the domain of energies and observation angles where the de-excited radiation dominates. The derivation of these conditions is obvious but rather awkward. We present only the final answer:
\begin{equation}\label{applicability}
\begin{gathered}
    a)\;2(\la k_-|\phi'|)^{2/3}\ll1,\qquad b)\;\frac{\la^{1/3}|\omega|}{2k_-^{2/3}|\phi'|^{5/3}}\ll1,\qquad c)\;\frac{k_-^{2/3}(\phi'')^2}{2(\la|\phi'|)^{1/3}|\omega|}\lesssim1,\\ d)\;\frac{1}{2\la|\omega|\ups^2_-(0)|\phi'|}\ll1,\qquad e)\;\frac{16\la^{1/3}\omega^2}{\pi\ups_-(0)k_-^{5/3}(\phi'')^2}\ll1.
\end{gathered}
\end{equation}
Here $a)$ is the condition that the last term in \eqref{jjbar} is much greater than the first one; $b)$ is the condition $h^2\ll1$; $c)$ is $B\lesssim1$ what is necessary since \eqref{spectr_dens_2} is exponentially suppressed for $B$ large; $d)$ is the radiation formation condition (the saddle points are located sufficiently far from the integration contour boundary); $e)$ implies that the contribution \eqref{spectr_dens_2} dominates over the contribution coming from the contour $C_0$. Graphically, these conditions are depicted in Fig. \ref{rad2_plots}.

Despite the fact that we consider the radiation of a particle that enters the electromagnetic field and then moves in it for an infinite time, let us estimate the formation time of the de-excited radiation. In terms of the variable \eqref{y_var}, the de-excited radiation is formed when
\begin{equation}
    y_f\gg\sqrt{B}.
\end{equation}
Hence, keeping in mind \eqref{applicability}, we obtain
\begin{equation}
    x_-^f\gg\Big(\frac{2|\phi'|}{\la|\omega|^3}\Big)^{1/2}.
\end{equation}
Eventually, employing formula (11) of \cite{BogKaz}, we come to (cf. (17) of \cite{BogKaz})
\begin{equation}
    |\phi'|\ll\frac{\la|\omega|}{2}\Big(\frac{6d}{\la}\Big)^{2/3},
\end{equation}
where the notation of \cite{BogKaz} was used.

\paragraph{Synchrotron entrance radiation.}

The ultrarelativistic particle entering the crossed electromagnetic field produce an intense hard electromagnetic radiation. This radiation resembles in many respects the ultrarelativistic limit of the synchrotron radiation. In particular, we shall see that the maximum of the spectral density of radiation is reached at the photon energy which is described quite well by the formula for the synchrotron radiation, and the most part of the radiation is concentrated in the cone opening $2/\gamma$. However, the form of the spectral angular distribution deviates from the ultrarelativistic limit of the synchrotron radiation due to the fact that the acceleration of a charged particle changes stepwise when it enters the electromagnetic field. This is not the edge radiation \cite{EdRadTh,SynchRad2015}, which is a long-wavelength radiation. Though, of course, the distortion of the synchrotron spectrum is the ``edge effect'', i.e., the result of the abrupt change of particle's acceleration. A similar radiation was discussed in \cite{BTFarc}.

\begin{figure}[tp]
\centering
\includegraphics*[width=0.4\linewidth]{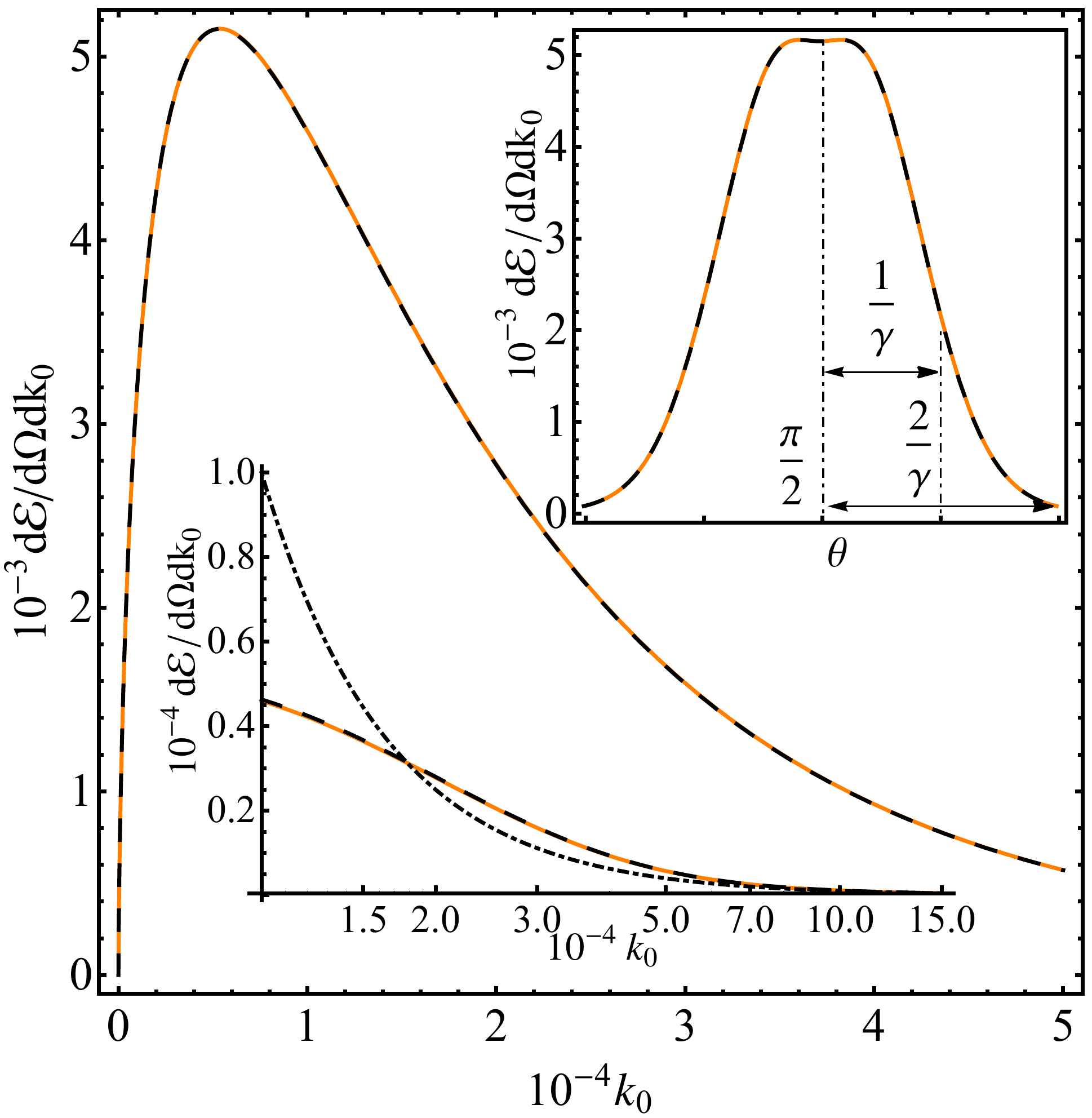}\;\;
\includegraphics*[width=0.415\linewidth]{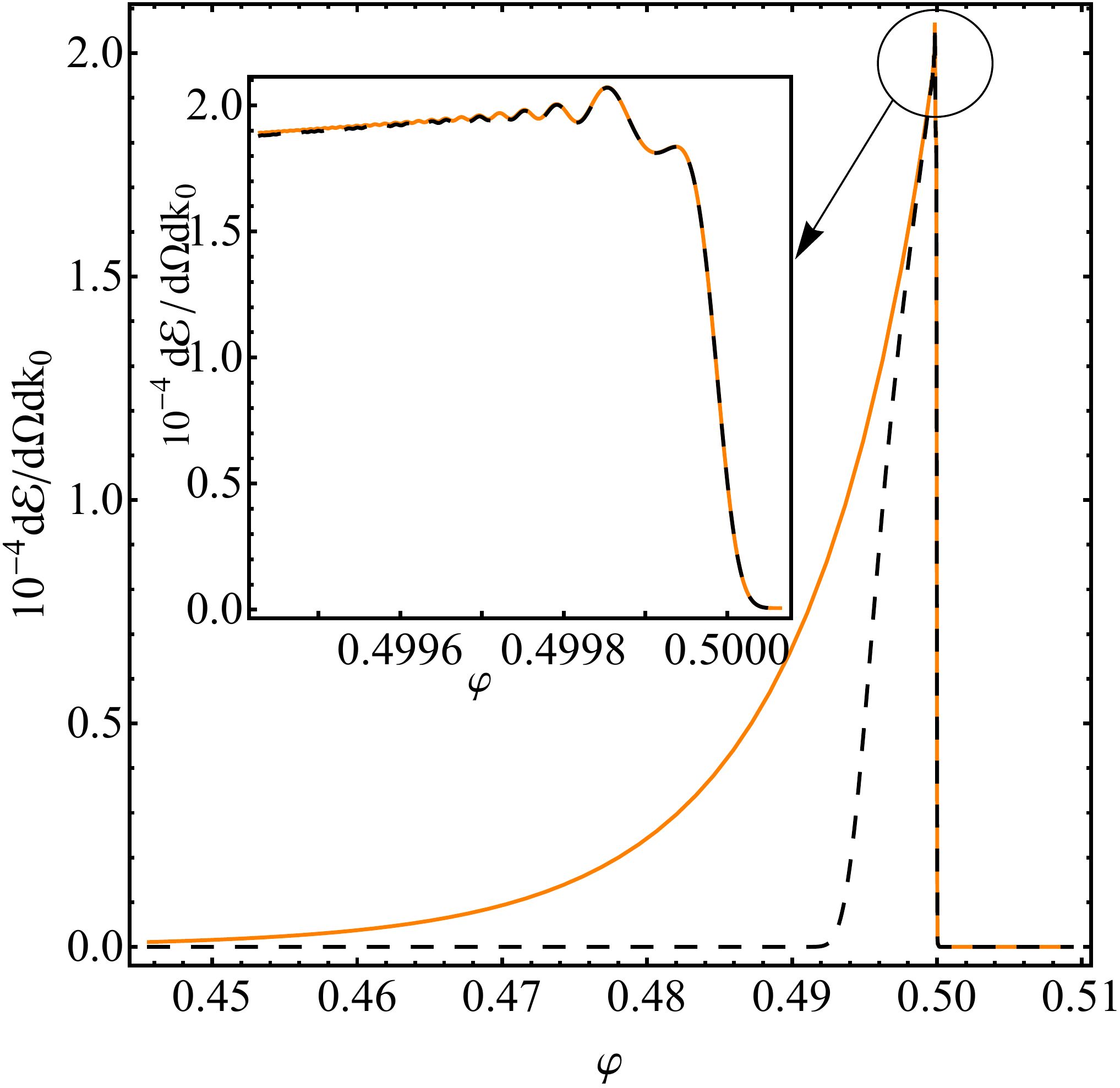}
\caption{{\footnotesize Synchrotron entrance radiation for $\ga=5\times10^3$, $\omega=-1.47\times10^{-4}$, $r_3(0)=0$, and the entrance angle $\pi/2$ counted from the $x^2$ axis. The orange line is the numerical result, the dashed line follows from \eqref{peak_first}, and the dot-dashed line is \eqref{spectr_dens_UV}. Left panel: The spectral angular distribution of the forward synchrotron entrance radiation. The inset: The spectral angular distribution of the synchrotron entrance radiation for $\vf=\pi/2$ and the photon energy $k_0^m\approx5358$  given by \eqref{ko_sinchrotr}. The small plot: The ultraviolet asymptotics of the spectral angular distribution. Right panel: The spectral angular distribution of the synchrotron entrance radiation for $\theta=\pi/2$ and the photon energy $k_0^m\approx5358$. The inset: The fine structure of the main peak. Formula \eqref{peak_first} leads to a spectral angular distribution describing with a good accuracy the real distribution in the cone opening less than or equal to $50\ga^{-1}$.}}
\label{rad1_plots}
\end{figure}

The derivation of the spectral angular distribution of radiation of an ultrarelativistic particle entering the electromagnetic field is analogous to the derivation of an ultrarelativistic asymptotics of the synchrotron radiation (see, e.g., \cite{LandLifshCTF.2,TerMikhKha,Bord.1}). It is convenient to parameterize particle's worldline by the laboratory time $t$. If the modulus of the radiation observation angle counted from the entrance direction of a particle is less than or of order $\ga^{-1}$, and the incoming particle is ultrarelativistic, then
\begin{equation}
    1-\be_\parallel\approx\frac{1+\ga^2\be_\perp^2}{2\ga^2},\qquad \ddot{\beta}_\parallel\approx-|\mathbf{a}_\perp|^2\approx -|\mathbf{a}|^2,\qquad \boldsymbol{\beta}_\perp:=\boldsymbol{\beta}-\beta_\parallel \mathbf{n},
\end{equation}
where $\mathbf{a}:=\dot{\boldsymbol{\be}}$, $\beta_\parallel:=(\boldsymbol{\be}\mathbf{n})$, and $\mathbf{n}:=\mathbf{k}/k_0$. Also
\begin{equation}
    k_\mu x^\mu\approx k_\mu x^\mu(0)+k_0\Big[(1-\be_\parallel)t-\dot{\beta}_\parallel\frac{t^2}{2}-\ddot{\beta}_\parallel\frac{t^3}{6}\Big],\qquad \boldsymbol{\be}_\perp\approx\boldsymbol{\be}_\perp(0)+\mathbf{a}_\perp(0)t.
\end{equation}
Therefore, up to an irrelevant common phase factor, the electric field strength of radiation created by a charged particle entering the electromagnetic field is
\begin{equation}\label{peak_first}
    R\mathbf{E}=e\frac{k_0\boldsymbol{\be}_\perp(0)}{k_\mu\dot{x}^\mu(0)}-iek_0\int_0^\infty dt\boldsymbol{\be}_\perp e^{ik_\mu[x^\mu(0)-x^\mu]}\approx e\Big\{\frac{\boldsymbol{\be}_{\perp}}{1-\be_\parallel}\big[1-iD\tilde{I}_0(x,D)\big]-i\frac{\mathbf{a}_\perp}{a_\perp}\frac{(2D)^{1/2}\tilde{I}_1(x,D)}{(1-\be_\parallel)^{1/2}}\Big\},
\end{equation}
where it is assumed that $\be_\perp\lesssim\ga^{-1}$, all the kinematic quantities after the approximate equality are taken at the initial instant of time, and the following notation was introduced:
\begin{equation}\label{InDx}
\begin{split}
    \tilde{I}_n(x,D):=\int_0^\infty dtt^ne^{-i(Dt+xt^2+t^3/3)},\qquad D&:=(1-\be_\parallel)\Big(\frac{2k^2_0}{a^2}\Big)^{1/3}\approx (1+\ga^2\be_\perp^2)\Big(\frac{k_0}{2a\ga^3}\Big)^{2/3},\\
    x&:=-\frac{a_\parallel}{2a}\Big(\frac{2D}{1-\be_\parallel}\Big)^{1/2}.
\end{split}
\end{equation}
The integration contour in $\tilde{I}_n(x,D)$ is shifted slightly below the real axis at $t\rightarrow+\infty$ so that the integral converges for all $\re n>-1$. Some properties of the functions $I_n(x,D)$ are presented in App. \ref{Prop_In}. The spectral angular distribution of radiation is obtained from \eqref{peak_first} by formula \eqref{spectr_dens}, the Stokes parameters being given by expressions \eqref{Stokes_par_gen}. The comparison with numerical results is presented in Fig. \ref{rad1_plots}.

For the forward radiation, $\boldsymbol{\be}_{\perp}=0$ and $x\approx0$. Then it follows from \eqref{peak_first} that the radiation is linearly polarized in this case for all the photon energies. The maximum of radiation power is reached at the photon energy
\begin{equation}\label{ko_sinchrotr}
    k^m_0=2D_{ext}^{3/2}a\gamma^3,\qquad D_{ext}\approx0.81,
\end{equation}
where $D_{ext}$ provides the maximum to the function $D|\tilde{I}_1(D)|^2$. Formula \eqref{ko_sinchrotr} is in a good agreement with the formula for the synchrotron radiation maximum (see, e.g., \cite{LandLifshCTF.2,TerMikhKha,Bord.1}).

\paragraph{Ultraviolet asymptotics.}

In order to find the ultraviolet asymptotics of radiation, one can use the WKB method for the boundary point in evaluating the integrals \eqref{current_fourier}. As we have already noted, the leading WKB contribution is canceled by the out of the integral term in \eqref{current_fourier}. The next to leading contribution gives
\begin{equation}
    d \mathcal{E}=-e^2\frac{\ddot{x}^2(k\dot{x})^2-2(\dot{x}\ddot{x})(k\dot{x})(k\ddot{x})+\dot{x}^2(k\ddot{x})^2}{(k\dot{x})^6}\frac{d\spk}{4\pi^2}\;\;\text{for}\;\; \Big|\frac{d}{d\tau}\frac{\dot{x}^\mu}{(k\dot{x})}\Big|\gg \Big|\frac{d}{d\tau}\frac{1}{(k\dot{x})}\frac{d}{d\tau}\frac{\dot{x}^\mu}{(k\dot{x})}\Big|,
\end{equation}
where the dot denotes the derivative with respect to an arbitrary time parameter, $4$-vectors are contracted with the aid of the Minkowski metric, and all the quantities are taken at the initial instant of time. Choosing the laboratory time as the time variable, we arrive at
\begin{equation}\label{spectr_dens_UV}
    d \mathcal{E}=e^2\frac{\mathbf{a}^2(1-\beta_\parallel)^2+2(\mathbf{a}\boldsymbol{\be})a_\parallel(1-\beta_\parallel) -a_\parallel^2(1-\be^2)}{(1-\beta_\parallel)^6}\frac{dk_0d\Omega}{4\pi^2 k_0^2},
\end{equation}
where $a_\parallel:=(\mathbf{a}\mathbf{n})$ and $\beta_\parallel:=(\boldsymbol{\beta}\mathbf{n})$. In increasing the radiated photon energy, the spectral density of radiation decreases as $k_0^{-2}$ \cite{BTFarc}. The applicability condition of the WKB expansion for the boundary point can be roughly written as
\begin{equation}
    \frac{a^2}{k_0^4(1-\be_\parallel)^4}\ll1.
\end{equation}
This radiation is linearly polarized with the polarization vector independent of the photon energy
\begin{equation}\label{polariz_vect_UV}
    \mathbf{e}=\frac{\mathbf{a}(1-\beta_\parallel)+(\boldsymbol{\beta}-\mathbf{n})a_\parallel}{|\mathbf{a}(1-\beta_\parallel)+(\boldsymbol{\beta}-\mathbf{n})a_\parallel|}.
\end{equation}
If $\mathbf{e}_\pi$ and $\mathbf{e}_\s$ are the polarization vectors of the $\pi$ and $\s$ components, respectively, then the Stokes parameters in this basis are
\begin{equation}\label{Stokes_par}
    \xi_1=2(\mathbf{e}_\pi\mathbf{e})(\mathbf{e}_\s\mathbf{e}),\qquad\xi_2=0,\qquad \xi_3=(\mathbf{e}_\pi\mathbf{e})^2-(\mathbf{e}_\s\mathbf{e})^2.
\end{equation}
The comparison of the ultraviolet asymptotics of the spectral angular distribution of radiation with its exact values is given in Fig. \ref{rad1_plots}.

\paragraph{Infrared limit.}

It is not difficult to find the infrared asymptotics $k_0\rightarrow0$ of the spectral angular distribution of radiation. The Fourier transform of the current density components in this case looks like
\begin{equation}
\begin{split}
    j_+&\approx-4i\sqrt{\frac{2}{\la}}|\omega|^{-3/2}(20\e\phi')^{-1/3}h^2\phi'^2+\frac{i\ups_+(0)}{k_\mu\dot{x}^\mu(0)},\\
    j_-&\approx\frac{\Gamma(1/5)}{5} \sqrt{\frac{2}{\la}}|\omega|^{-3/2}(20\e\phi')^{-1/3} \Big(\frac{20}{ih^2}\Big)^{1/5}+\frac{i\ups_-(0)}{k_\mu\dot{x}^\mu(0)},\\
    j_3&\approx r_3(0)\frac{\Gamma(1/5)}{5} \sqrt{\frac{2}{\la}}|\omega|^{-3/2}(20\e\phi')^{-1/3} \Big(\frac{20}{ih^2}\Big)^{1/5}+\frac{i\ups_3(0)}{k_\mu\dot{x}^\mu(0)},\\
    j_1&\approx-\sgn(\omega)\frac{\Ga(3/5)}{5} \sqrt{\frac{2}{\la}}|\omega|^{-3/2}(20\e\phi')^{-1/3}h^2\phi'\Big(\frac{20}{ih^2}\Big)^{3/5}+\frac{i\ups_1(0)}{k_\mu\dot{x}^\mu(0)}.
\end{split}
\end{equation}
Consequently, in the leading order at $k_0\rightarrow0$, we have
\begin{equation}
    j_\mu^*j^\mu \approx\frac{k_-\dot{x}^2-2(k\dot{x})\ups_-}{k_-(k\dot{x})^2}=\frac{(k_-\dot{x}_\mu-e^-_\mu(k\dot{x}))^2}{k_-^2(k\dot{x})^2},
\end{equation}
where all the quantities on the right-hand side are taken at the initial instant of time. It is also easy to check that, in this approximation,
\begin{equation}
    j_\mu j^\mu=-j_\mu^*j^\mu,
\end{equation}
i.e., the radiation is linearly polarized. The spectral angular distribution of radiation and the polarization vectors are written as
\begin{equation}\label{spectr_dens_0}
    d \mathcal{E}=e^2\frac{2(1-\be_\parallel)(1-(\boldsymbol{\be}\mathbf{e}_-))-(1-e_-^\parallel)(1-\be^2)}{(1-e_-^\parallel)(1-\be_\parallel)^2}\frac{dk_0 d\Omega}{4\pi^2},\qquad \mathbf{e}=\frac{\boldsymbol{\be}(1-e_-^\parallel)-\mathbf{e}_-(1-\be_\parallel)-\mathbf{n}(\be_\parallel-e_-^\parallel)}{|\boldsymbol{\be}(1-e_-^\parallel)-\mathbf{e}_-(1-\be_\parallel)-\mathbf{n}(\be_\parallel-e_-^\parallel)|},
\end{equation}
where $e_-^\parallel:=(\mathbf{e}_-\mathbf{n})$. The Stokes parameters are found according to \eqref{Stokes_par}. Notice that formula \eqref{spectr_dens_0} is exact in the limit $k_0\rightarrow0$.

\section{Conclusion}

We have described in detail both analytically and numerically the properties of radiation created by a classical scalar charged particle in a crossed field with radiation reaction taken into account employing the exact solution to the Landau-Lifshitz equation. The results of numerical simulations illustrating and confirming the analytical formulas are summarized in Figs. \ref{rad2_plots}, \ref{rad1_plots}. These results clearly show the existence of the two types of electromagnetic radiation, which we call the de-excited and synchrotron entrance ones. These two types of radiation possess distinct characteristic energy scales, polarization properties, and angles of observation. All these characteristics were described in the paper.

\paragraph{Acknowledgments.}

We appreciate V.G. Bagrov and D.V. Karlovets for fruitful conversations. The work is supported by the RFBR grants No. 16-02-00284 and No. 16-32-00464-mol-a, and by the Russian Federation President grant No. MK 5202.2015.2.

\appendix
\section{Asymptotics of the function $F(\phi)$}\label{F_phi_Asymptotes}

Let us consider the integral
\begin{equation}
    \int_C dy e^{-i(y^5+a y^3+by)},\quad a,b\in \mathbb{C},
\end{equation}
where the integration contour $C$ is depicted in Fig. \ref{cont_airy}. This integral is an entire function of $a$ and $b$. We are interested in the case when
\begin{equation}
    a=-\frac{10}{3}\phi',\qquad b=5|\phi|^2,\qquad \phi=\phi'+i\phi'',\quad \phi',\phi''\in \mathbb{R}.
\end{equation}
Let us define
\begin{equation}
    F(\phi):=\int_C dy e^{-i(y^5-\frac{10}{3}\phi' y^3+5|\phi|^2y)}.
\end{equation}
We give the expansion of the function $F(\vf)$ in the following regions
\begin{equation}\label{fphi_cases}
    (i)\,|\phi'|^{5/6}\gg20^{-1/3},\,|\phi''|^4\ll4|\phi'|^{3/2}, \qquad (ii)\,25|\phi''|^{3/2}\gg|\phi|^{1/4}, \qquad(iii)\,|\phi|\ll1.
\end{equation}
In the case (i), we have the asymptotic expansion
\begin{multline}\label{Airy_appr}
    F(\phi)=\frac{2\pi}{(20\phi')^{1/3}}e^{-i[\frac83(\phi')^{5/2}+5(\phi'')^2(\phi')^{1/2}]}\Big\{\Ai(B)-\\
    -\frac{ih}{4}\big(B^2\Ai(B)+2\Ai'(B)\big)-h^2\Big[\Big(\frac{B^4}{32}+\frac{27B}{40}\Big)\Ai(B) +\frac{13B^2}{40}\Ai'(B)\Big]+\cdots \Big\},
\end{multline}
where $h:=20^{-1/3}(\vf')^{-5/6}$, $B:=5(\phi'')^2(20\phi')^{-1/3}$. The condition (i) means that
\begin{equation}\label{asympt_cond1}
    h\ll1,\qquad B^2h\ll1.
\end{equation}
For $\phi'<0$, it is necessary to put $\phi'\rightarrow\phi'-i0$ in \eqref{Airy_appr} and take the principal branches of the multi-valued functions.

In the case (ii), the standard WKB method gives in the leading order
\begin{equation}
    F(\phi)\approx\sqrt{\frac{\pi}{10\phi''\sqrt{\phi^*}}}e^{-i\frac{10}{3}\sqrt{\phi^*}\big[|\phi|^2-\frac15(\phi^*)^2\big]}.
\end{equation}
The condition (ii) guarantees the applicability of the WKB method. In the case (iii), we deduce
\begin{multline}
    F(\phi)=(i+e^{-i\frac{\pi}{10}})\Ga\Big(\frac65\Big)-\frac23(i-e^{i\frac{\pi}{10}})\Ga\Big(\frac45\Big)\phi'-(i-e^{-i\frac{7\pi}{10}}) \Ga\Big(\frac25\Big)\Big[|\phi|^2-\frac49(\phi')^2\Big]+\\
    +\frac52(i+e^{i\frac{7\pi}{10}})\Ga\Big(\frac35\Big)\Big[|\phi|^4-\frac43(\phi')^2|\phi|^2+\frac{32}{81}(\phi')^4\Big]+o(|\phi|^4).
\end{multline}

\section{Functions $\tilde{I}_n(x,D)$}\label{Prop_In}

The function $\tilde{I}_n(x,D)$ (see \eqref{InDx}) is an entire function of the complex variables $x$ and $D$. The following representation holds
\begin{equation}\label{IxD_expan}
    \tilde{I}_0(x,D)=\sum_{n=0}^\infty(-ix)^{n}\frac{\tilde{I}_{2n}(D)}{n!},\qquad \tilde{I}_n(D):=\tilde{I}_n(0,D).
\end{equation}
An analogous expansion is valid for $\tilde{I}_1(x,D)$ with the replacement of $\tilde{I}_{2n}(D)$ by $\tilde{I}_{2n+1}(D)$. In virtue of the relations,
\begin{equation}
    \tilde{I}''_0(D)=D\tilde{I}_0(D)+i,\qquad \tilde{I}_n(D)=i^n\tilde{I}^{(n)}_0(D),
\end{equation}
all $\tilde{I}_n(D)$, $n\in \mathbb{N}$, are expressed through $\tilde{I}_0(D)$ and $\tilde{I}_1(D)$. For example,
\begin{equation}
\begin{alignedat}{2}
    \tilde{I}_2(D)&=-D\tilde{I}_0(D)-i,&\qquad \tilde{I}_3(D)&=-D\tilde{I}_1(D)-i\tilde{I}_0(D),\\
    \tilde{I}_4(D)&=D^2\tilde{I}_0(D)-2i\tilde{I}_1(D)+iD,&\qquad \tilde{I}_5(D)&=D^2\tilde{I}_1(D)+4iD\tilde{I}_0(D)-3,\\
    \tilde{I}_6(D)&=-(D^3+4)\tilde{I}_0(D)+6iD\tilde{I}_1(D)-iD^2,&\qquad \tilde{I}_7(D)&=-(D^3+10)\tilde{I}_1(D)-9iD^2\tilde{I}_0(D)+8D.
\end{alignedat}
\end{equation}
The functions $\tilde{I}_0(D)$, $\tilde{I}_1(D)$ are reduced to the known special functions
\begin{equation}
\begin{split}
    \tilde{I}_0(D)&=\pi\Ai(D)-\frac{i\pi}3\Bi(D)+\frac{iD^2}{2}{}_1F_2\big(1;\tfrac43,\tfrac53;\tfrac{D^3}{9}\big),\\
    \tilde{I}_1(D)&=i\pi\Ai'(D)+\frac{\pi}3\Bi'(D)-D{}_1F_2\big(1;\tfrac43,\tfrac53;\tfrac{D^3}{9}\big)-\frac{3D^4}{40}{}_1F_2\big(2;\tfrac73,\tfrac83;\tfrac{D^3}{9}\big).
\end{split}
\end{equation}
The function $\tilde{I}_0(D)$ (up to the common factor $\pi$) is also known as the Scorer function \cite{NIST}. The hypergeometric function entering into $\tilde{I}_0(D)$ is an entire function of $D$ and can be expressed through the Anger functions \cite{PruBryMarIII}
\begin{equation}
    {}_1F_2\big(1;\tfrac43,\tfrac53;\tfrac{z^3}{9}\big)=\frac{4\pi}{3\sqrt{-3z^3}}\Big[\mathbf{J}_{1/3}\big(\tfrac23\sqrt{-z^3}\big)-\mathbf{J}_{-1/3}\big(\tfrac23\sqrt{-z^3}\big) \Big].
\end{equation}
In the ultrarelativistic limit and for $\be_\perp\lesssim\ga^{-1}$, the magnitude of $x$ is rather small (see \eqref{InDx}). For $|x|\ll1$, the expansion \eqref{IxD_expan} is rapidly converging and so one may keep only a few first terms in it. In the case $|x|\lesssim1$, it is useful to employ the other representation of the functions $\tilde{I}_0(x,D)$ and $\tilde{I}_1(x,D)$:
\begin{equation}
\begin{split}
    \tilde{I}_0(x,D)&=e^{ix(D-2x^2/3)}\Big[\tilde{I}_0(D-x^2)+e^{-\frac13\frac{\partial^3}{\partial D^3}}\frac{e^{-ix(D-x^2)}-1}{i(D-x^2)}\Big],\\
    \tilde{I}_1(x,D)&=e^{ix(D-2x^2/3)}\Big[\tilde{I}_1(D-x^2)+e^{-\frac13\frac{\partial^3}{\partial D^3}}\frac{\partial}{\partial D}\frac{e^{-ix(D-x^2)}-1}{D-x^2}\Big]-x\tilde{I}_0(x,D).
\end{split}
\end{equation}
If $|x|\lesssim1$ then the expansion of $\exp[-\tfrac13\partial^3/\partial D^3]$ can be terminated with a few first terms left.

The function $\tilde{I}_0(x,D)$ as a function of $D$ satisfies the linear inhomogeneous ordinary differential equation
\begin{equation}
    \tilde{I}''_0(x,D)-2ix\tilde{I}'_0(x,D)-D\tilde{I}_0(x,D)=i,
\end{equation}
where the prime denotes the derivative with respect to $D$. After the substitution $\tilde{I}_0(x,D)=e^{ixD}f(D-x^2)$, this equation is reduced to the inhomogeneous Airy equation for the function $f(t)$. The solution of this equation is found by the standard means. As a result, we deduce the integral representations:
\begin{multline}
    \tilde{I}_0(x,D)=\pi c_1(x) e^{ixD}\Ai(D-x^2)-\pi c_2(x) e^{ixD}\Bi(D-x^2)\\
    +i\pi\Bi(D-x^2)\int_0^D dse^{ix(D-s)}\Ai(s-x^2)-i\pi\Ai(D-x^2)\int_0^D dse^{ix(D-s)}\Bi(s-x^2),
\end{multline}
where
\begin{equation}
\begin{split}
    c_1&=e^{-2ix^3/3}-x\big[\big(\Bi'(-x^2)+ix\Bi(-x^2)\big){}_2F_2(\tfrac12,1;\tfrac23,\tfrac43;-\tfrac{4ix^3}{3}) -\tfrac{3ix}{2}\Bi(-x^2){}_2F_2(1,\tfrac32;\tfrac43,\tfrac53;-\tfrac{4ix^3}{3})\big],\\
    c_2&=i\frac{e^{-2ix^3/3}}{3}-x\big[\big(\Ai'(-x^2)+ix\Ai(-x^2)\big){}_2F_2(\tfrac12,1;\tfrac23,\tfrac43;-\tfrac{4ix^3}{3}) -\tfrac{3ix}{2}\Ai(-x^2){}_2F_2(1,\tfrac32;\tfrac43,\tfrac53;-\tfrac{4ix^3}{3})\big],
\end{split}
\end{equation}
and also
\begin{multline}
    \tilde{I}_0(x,D)=d_1(x) e^{ix(D-x^2)+i\pi/6}\Ai(D-x^2)-d_2(x) e^{ix(D-x^2)+i\pi/6}\Bi(D-x^2)\\
    +i\pi\Bi(D-x^2)\int_0^{D-x^2} dse^{ix(D-x^2-s)}\Ai(s)-i\pi\Ai(D-x^2)\int_0^{D-x^2} dse^{ix(D-x^2-s)}\Bi(s),
\end{multline}
where
\begin{equation}
    d_1=\frac{e^{\frac{ix^3}{3}}}2\big[\Ga(\tfrac13)\Ga(\tfrac23,\tfrac{ix^3}{3})-e^{2\pi i/3}\Ga(\tfrac23)\Ga(\tfrac13,\tfrac{ix^3}{3})\big],\qquad
    d_2=\frac{e^{\frac{ix^3}{3}}}{2\sqrt{3}}\big[\Ga(\tfrac13)\Ga(\tfrac23,\tfrac{ix^3}{3})+e^{2\pi i/3}\Ga(\tfrac23)\Ga(\tfrac13,\tfrac{ix^3}{3})\big].
\end{equation}

\end{document}